\documentclass[%
 reprint,
superscriptaddress,
 amsmath,amssymb,
onecolumn
]{revtex4-1}

\usepackage{graphicx}
\usepackage{dcolumn}
\usepackage{bm}
\usepackage{xcolor}
 \usepackage[mathlines]{lineno}

\begin{document}

\preprint{APS/123-QED}

\title{Waves beneath a drop levitating over a moving wall}



\author{Kyle I. McKee}
\author{Bauyrzhan K. Primkulov}%
\affiliation{%
 Department of Mathematics, Massachusetts Institute of Technology,\\
 Cambridge, MA 02139, USA
}%

\author{Kotaro Hashimoto}
\author{Yoshiyuki Tagawa}
\affiliation{
 Department of Mechanical Systems Engineering, Tokyo University of Agriculture and Technology,\\
 Tokyo, Japan
}%
\author{John W.M. Bush}%
\email{bush@math.mit.edu}
\affiliation{%
 Department of Mathematics, Massachusetts Institute of Technology,\\
 Cambridge, MA 02139, USA
}%

\date{August 21, 2024;\;Accepted for publication in \emph{Physical Review Fluids}}

\begin{abstract}
In recent experiments, \citet{sawaguchi2019droplet} directly probed the lubrication layer of air beneath a droplet levitating inside a rotating cylindrical drum. For small rotation rates of the drum, the lubrication film beneath the drop adopted a steady shape, while at higher rotation rates, travelling waves propagated along the drop's lower surface with roughly half the wall velocity. 
We here rationalize the physical origin of these waves. We begin with a simplified model of the lubrication flow beneath the droplet, and examine the linear stability of this base state to perturbations of the Tollmien--Schlichting type. Our developments lead to the Orr-Sommerfeld equation (OSE), whose eigenvalues give the growth rates and phase speeds of the perturbations. By considering wavelengths long relative to the lubrication film thickness, we solve the OSE perturbatively and so deduce the wavelength and phase velocity of the most unstable mode. We find satisfactory agreement between experiment and theory over the parameter regime considered in the laboratory.
\end{abstract}

\maketitle


\section{\label{sec:level1}Introduction}


Sheared interfaces are ubiquitous in nature and technology, 
arising when wind blows over oceans and when liquid films flow down inclined planes \cite{caulfield2021layering,balmforth2004dynamics,weinstein2004coating}. The question of the stability of such flows has thus received considerable attention \cite{caulfield2021layering,oron1997long, drazin2004hydrodynamic}. 
The classic Kelvin-Helmholtz instability rationalizes the waves generated by shear across an interface of two inviscid fluid layers, and has been studied extensively in the context of wind-driven flow over water~\citep{ursell1956wave,phillips1957generation}. Alternatively, for gas flow over a thin liquid film, the film stability may be assessed via the lubrication approximation wherein viscous effects becomes significant \citep{craik1966wind,miles1960hydrodynamic,oron1997long,oron1997long}. For example, \citet{kapitza1948wave} showed that when gravity drives a liquid film down an incline, capillary waves may emerge. The stability of thin viscous film flows has been extensively explored by the fluid mechanics community~\cite{benjamin1957wave,yih1963stability,oron1997long,Craster-RMP}. \\

The approach of two liquid phases, or that of a liquid towards a solid, is resisted by lubrication pressures generated in the intervening fluid~\cite{kavehpour2015coalescence}.
For example, sheared air films extend the lifetime of milk drops levitating on coffee \citep{geri2017thermal}, may suppress coalescence of liquid jets impinging on a liquid bath \cite{wad,thrasher2007bouncing}, and may preclude contact of droplets impacting on a wet incline \cite{gilet2012droplets} or a solid substrate \cite{liu2015kelvin,Kolinski_2014}. 
In the present paper, we examine the lubrication flow beneath a levitating, rolling droplet. We first describe the mechanics of such drops, then assess the stability of the underlying air layer.

\begin{figure}
\centering
 \includegraphics[width=.95\linewidth]{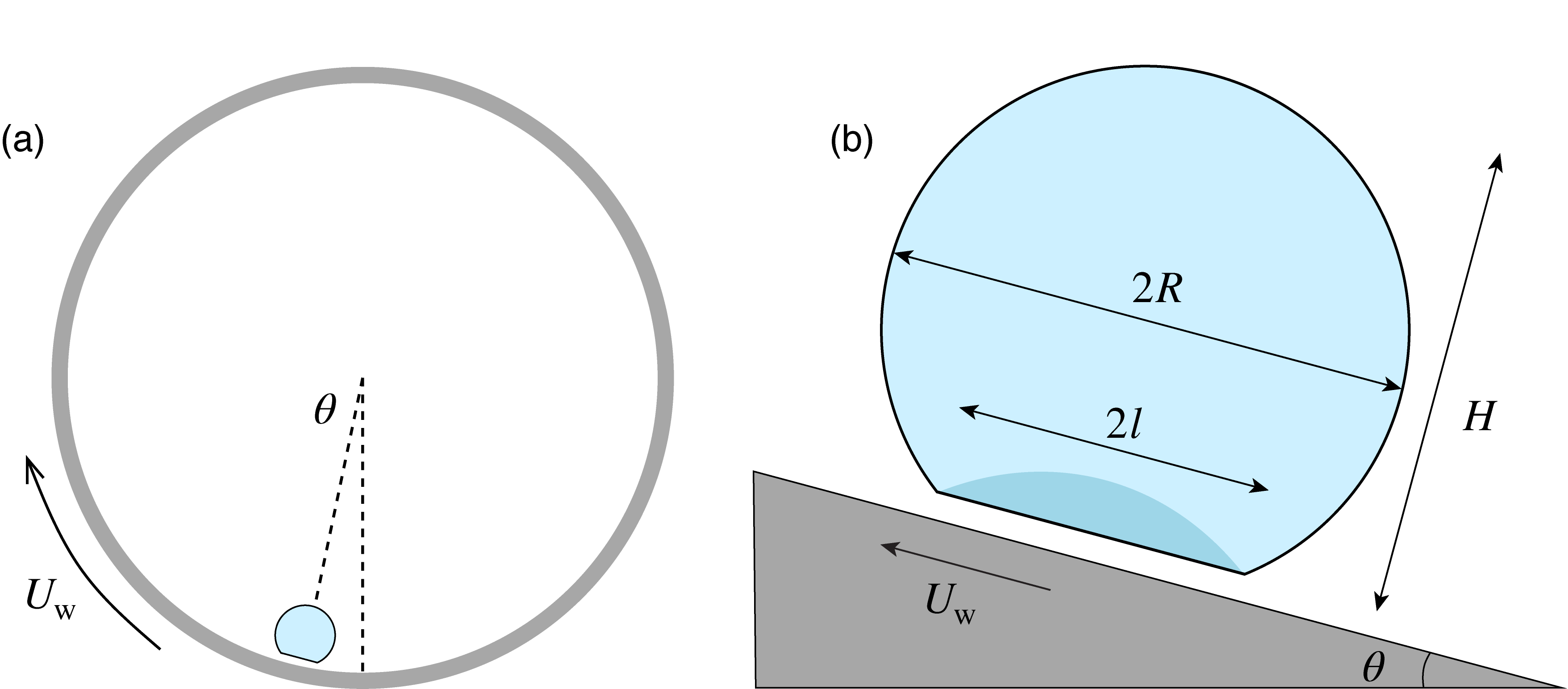}
 \caption{\label{fig:fig_schem} (a)~Schematic of a droplet rolling in a rotating drum. (b)~The shape of the droplet is determined by the balance of surface tension and gravity. Viscous drops advance via rigid body apart from the shear experienced in the highlighted dark region\citep{mahadevan1999rolling,aussillous2001liquid,aussillous2006properties}. In the regime considered here, the drop deformation is characterized by height-to-radius ratios in the range $0.85<H/R<1.25$, where $H/R=2$ corresponds to the case of an undeformed sphere. \label{fig:dropgeom}}
\end{figure}

\color{black}
It is well-known that a droplet may be suspended indefinitely on a thin layer of its own vapour when placed above a hot substrate — the so-called Leidenfrost effect  \citep{leidenfrost1756aquae,quere2013leidenfrost}. However, it is also possible to sustain a droplet above a thin lubrication flow in an isothermal setting. Specifically, relative motion between the drop and its substrate may induce lubrication pressures in the gas flow beneath the drop, thereby providing a lift force that supports its weight. For example, \citet{sreenivas1999levitation} achieved stable isothermal levitation of a drop on the free surface of a hydraulic jump and \citet{lhuissier2013levitation} levitated a drop inside of a cylinder rotating about a horizontal axis (see figure \ref{fig:dropgeom}). \citet{gauthier2016aerodynamic} studied droplets levitating atop a table rotating about a vertical axis, and coined the phrase ``aerodynamic Leidenfrost effect'' to describe this dynamic levitation.

\citet{sawaguchi2019droplet} revisited the rotating cylinder experiments of \citet{lhuissier2013levitation} and investigated the details of the interface shape and pressure distribution beneath levitating drops of 50-5000 cSt Si oil with a typical diameter  of $5\mathrm{mm}$. When the cylinder wall velocity beneath the droplet was just large enough to support levitation, a stable interface was observed below the drop. However, when the wall velocity was increased, unsteady travelling waves were observed along the base of the drop. Typically, these waves had a phase velocity of roughly half the wall velocity and a wavelength on the order of 1-2 mm. Compared to the thickness of the air film, $h_1 \sim 10 \mu \mathrm{m}$, these are \emph{long} waves, a fact that will be exploited in our theoretical developments. While \citet{sawaguchi2019droplet} mentioned the emergence of such waves, and \citet{yoshi} suggested that they may be shear driven, these waves have yet to be rationalized theoretically. The objective of the current paper is to do so.

After developing a simple model for the base flow beneath the drop, we build on the analysis of \citet{yih1967instability} to assess the stability of long wavelength perturbations of the Tollmien-Schlichting type. We note that neither inviscid nor viscous potential flow models \citep{funada2001viscous} capture the waves of interest (see Appendix \ref{Inv_Visc_pot_flos}). 
Moreover, the frequency of the observed waves is markedly difference from the natural vibration frequency of the droplet (see Appendix \ref{resjust}); thus, the observed waves can not simply be rationalized in terms of advection of the droplet's natural modes of vibration. 

The remainder of this paper is arranged as follows. 
In \S \ref{physicalpic}, we present a physical picture of the levitating droplet system of interest.  In \S \ref{mathform}, we lay out the framework for analysing the stability of the lubrication flow beneath the droplet. In \S \ref{perturb}, we solve the equations of \S \ref{mathform} perturbatively via a long wavelength expansion, and predict a fastest growing wavelength and its corresponding phase velocity. In \S \ref{detd2}, we compare the predictions of our model with experiments. In \S \ref{disc}, we discuss the limitations of our model and discuss other settings where similar instabilities might arise.

\section{Experimental Procedure and Physical Picture}\label{physicalpic}
\begin{figure}
\centering
 \includegraphics[width=\linewidth]{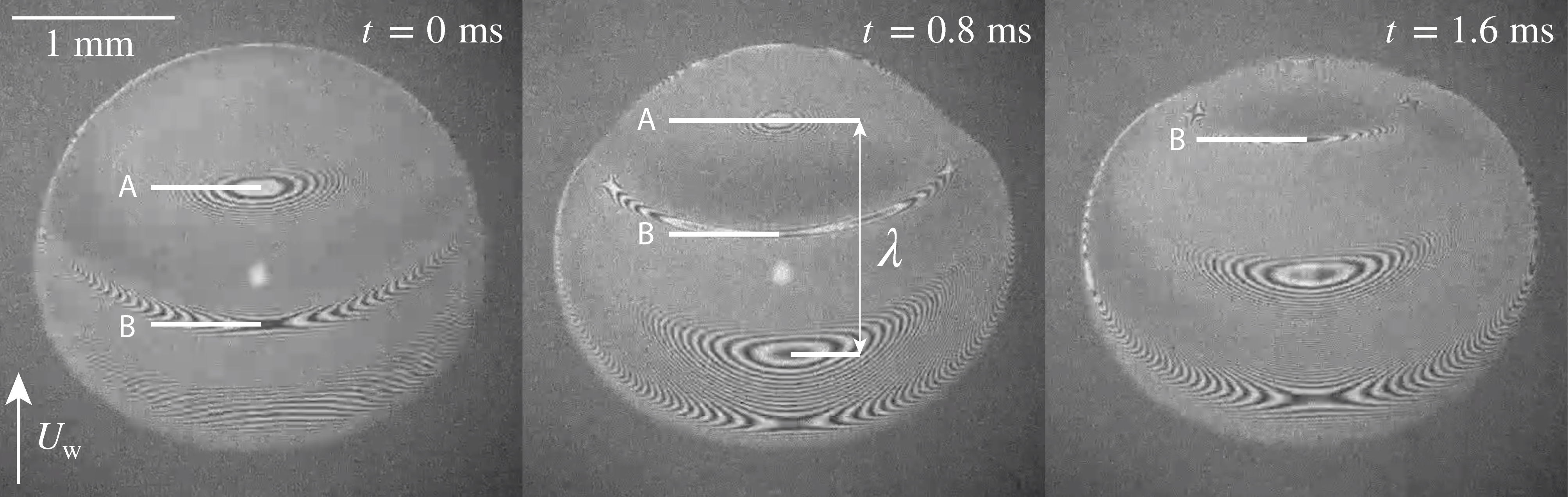}
 \caption{\label{fig:fig_exp_im} Experimental snapshots of the wave progression beneath the rolling droplet. These images are taken from below using interferometry, for an experiment where $U=2\mathrm{m/s}$. $A$ and $B$ track the instability extrema between frames.}
\end{figure}

We consider the experimental arrangement employed by \citet{lhuissier2013levitation}, \citet{sawaguchi2019droplet}, and \citet{yoshi} depicted in figure \ref{fig:dropgeom}. We placed silicone oil drops (100 cSt, 20.9 mN/m) with radii $R$ ranging between 1.7 and 3.4 mm on the inner surface of a hollow silica-glass cylinder rotating on its axis of symmetry. The inner diameter of the cylinder was 200 mm and its angular velocity was used to control the wall speed $U_\mathrm{w}$, which ranged between 1.05 m/s and 2.10 m/s. In this parameter regime, the droplet rests on an air lubrication layer  and equilibrates to a static angle $\theta$ on the inner surface of the rotating cylinder. At this relatively large droplet viscosity, the bulk of the drop is in a rigid-body rotation except in a neighborhood adjoining the base of the drop  \citep[see Supplementary Movie 2]{sawaguchi2019droplet}.
At sufficiently high cylinder rotation rates, the droplet may develop travelling surface waves on its base~\citep[]{yoshi}. These surface waves were imaged with a high-speed camera (FASTCAM SA-X, Photron Co.) placed outside the rotating cylinder and directed towards the base of the drop. A beam of 630 nm monochromatic light was directed from the camera towards the base of the drop using a coaxial zoom lens (12x Co-axial Ultra Zoom Lens, Navitar Co.). The reflected light from the base of the drop then formed interference fringes, as shown in Fig.~\ref{fig:fig_exp_im}. High-speed imaging of the interference fringes allowed us to measure both wavelength and the wave speed of the surface waves.

The pressure, $p_{\mathrm{lub}}$, induced by the lubrication flow in a thin air layer of thickness $d_1 \approx 10\mu \mathrm{m}$ supports the drop's weight, preventing contact with the substrate. The force balance normal to the cylinder surface requires, 
\begin{equation}\label{eq:norm}
\rho g V \cos{\left(\theta\right)} \sim \pi l^2 p_{\mathrm{lub}},
\end{equation}
where $V$ is the droplet volume and $\pi l^2$ is the drop's contact area. 

The angle $\theta$ is set by a balance between gravity, drag, and viscous shear forces on the drop \citep{lhuissier2013levitation}. The tangential force balances reads,
\begin{equation}\label{eq:tang}
    \rho g V \sin\left(\theta\right)  = F_{\mathrm{drag}} + \pi l^2\tau_{\mathrm{lub}},
\end{equation}
where $\tau_{\mathrm{lub}}$ is the viscous stress in the lubrication layer. Experimentally, the majority of drops sit at an angle not exceeding 15 degrees. The approximation $\cos{\left(\theta\right)}\approx 1$ introduces an error of less than 5\% so that the normal force balance (\ref{eq:norm}) can be solved independent of (\ref{eq:tang}) and the equations become effectively decoupled. According to this approximation, since $p_{\mathrm{lub}}$ depends on $U_\mathrm{w}$ and $d_1$, $d_1$ is independent of angle. Since our model will only require $d_1$, we obviate the need to solve the tangential force balance, (\ref{eq:tang}). As the wall speed, $U_\mathrm{w}$, is increased beyond a critical threshold, traveling waves with wavelengths $\lambda \approx 1 \mathrm{mm}$ appear at the lower droplet interface and move at a speed of roughly $U_\mathrm{w}/2$ in the direction of the wall motion. Figure \ref{fig:fig_exp_im} displays interferometry images of these waves.

\begin{figure}
\centering
 \includegraphics[width=0.9\linewidth]{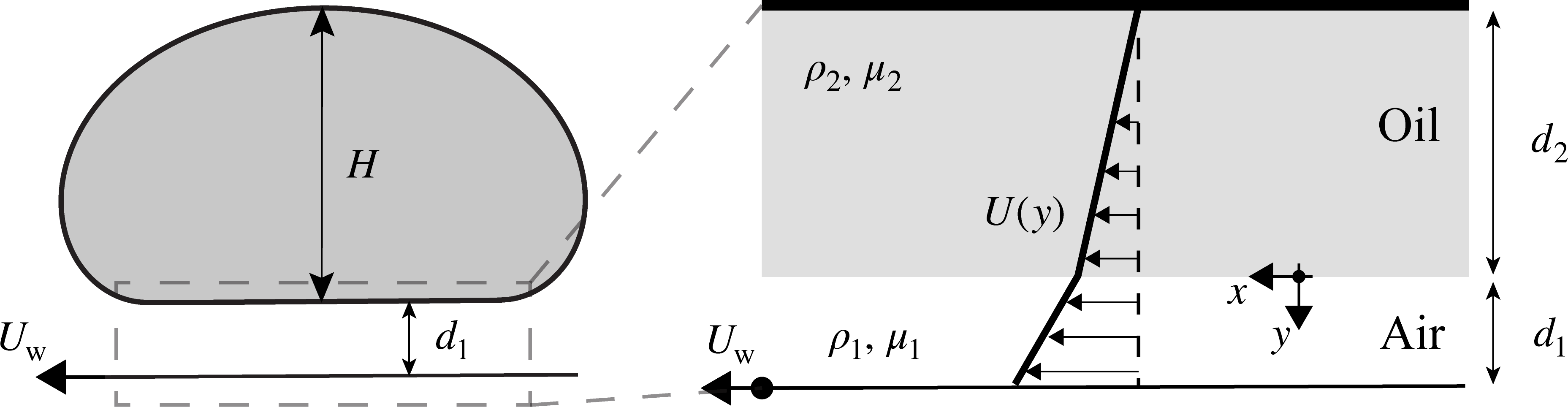}
 \caption{\label{fig:FlowSchem} Simplified model of the lubrication flow beneath the droplet. A linear Couette flow is assumed to arise in both the oil layer of thickness $d_2$, and the air lubrication layer of thickness $d_1$. While $d_1$ is taken as the average lubrication layer beneath the droplet (see \S \ref{physicalpic}), $d_2$ is the average shear thickness in the drop which is determined by a single fitting parameter across all experiments (see \S \ref{detd2}). }
\end{figure}

To simplify the flow geometry, we approximate the interface between the drop and the air film as flat (see figure \ref{fig:FlowSchem}). The lubrication flow is then modelled by a planar bounded Couette flow with the bottom wall moving at the cylinder speed, $U_\mathrm{w}$, and an upper effective wall, a distance $d_2$ inside the oil layer, where the fluid comes to rest. In their experiments, \citet[pp. 273, figure 12]{sawaguchi2019droplet} measured the pressure beneath the droplet, demonstrating its constancy away from a small dimple region formed at the edge of the drop. This constancy gives rise to a linear velocity profile within the lubrication layer, since lubrication analysis dictates that quadratic components of velocity only arise in the presence of pressure gradients.

We proceed by outlining how the lubrication film height, $d_1$, and the effective oil shear layer thickness, $d_2$, are determined. Since $d_1$ is independent of $\theta$ for small $\theta$, the situation may be related to the flat configuration of \citet{gauthier2016aerodynamic}. In their experiments, they found that the lubrication layer height scales as $d_1=C R\mathrm{Ca}^{2/3}$, with $C=0.6 \pm 0.2$, when $\mathrm{Ca}_{\mathrm{air}}\ll 1$. Note that the Capillary number in our experiments is small, $\mathrm{Ca}_{\mathrm{air}}=\mu_1 U_\mathrm{w}/\sigma\sim 10^{-3}$, suggesting that their scaling should also apply in our setting. When considering small drops ($R \ll l_c$) that are nearly spherical, so that $H \approx 2R$, the scaling of \citet{gauthier2016aerodynamic} becomes $d_1=C (H/2)\mathrm{Ca}^{2/3}$. Conversely, large tank-treading droplets ($R \gg l_c$) at low Reynolds number conform to the scaling $d_1 \sim l_c \mathrm{Ca}^{2/3}$ \cite{hodges2004sliding}. Noting that the height of such large drops saturates to $H=2l_c$, the scaling recast in terms of height becomes $d_1\sim H \mathrm{Ca}^{2/3}$.
We thus adopt the following scaling for our problem,
\begin{equation}\label{eq:d1eq}
    d_1=C_H H\mathrm{Ca}^{2/3},
\end{equation}
which scales appropriately in both the large and small $\mathrm{Bo}$ limits. In the small drop limit of (\ref{eq:d1eq}), the experiments of \citet{gauthier2016aerodynamic} indicate that $C_H=0.3\pm 0.1$.

What remains is to define $d_2$, the effective shear layer thickness at the base of the drop. Small drops ($\mathrm{Bo}\ll 1$) at low Capillary and Reynolds numbers reach a state of nearly solid-body rotation, and dissipation is confined to a neighborhood adjoining the contact area with vertical extent $d_2 \sim \mathrm{Bo}^{1/2} R$ (see Eq. 3 in \citep[]{mahadevan1999rolling}). 
For large tank-treading drops, \citet[see figure 5(b)]{hodges2004sliding} found that the shear layer permeates the entirety of the drop, so $d_2\sim H$.

The drops considered in the present paper have $\mathrm{Bo}=\mathcal{O}(1)$, lying awkwardly between the cases analysed by \citet{mahadevan1999rolling} and \citet{hodges2004sliding}. We thus expect a scaling for $d_2$ that is bounded by these two extreme cases. For the sake of simplicity, we introduce the simple scaling law, $d_2=AH$, where $A$ is a fitting parameter to model the shear layer thickness. For small drops, the scaling of \citet{mahadevan1999rolling}, $d_2\sim \mathrm{Bo}^{1/2} R$, leads to $A=\mathrm{Bo}^{1/2}/2$ so that $A \ll 1$. For large drops, the scaling of \citet{hodges2004sliding}, $d_2 \sim H$, implies that $A=\mathcal{O}(1)$. For our drops, we thus expect $A$ to be less than $1/2$. We discuss the fitting of $A$ in \S \ref{detd2}, where we deduce that $A=0.08$ adequately describes our experiments.


\section{Stability Analysis}




\subsection{Mathematical formulation and stability analysis}\label{mathform}
We begin by writing down the base flow as depicted in figure \ref{fig:FlowSchem}. Let us first define non-dimensional quantities to be used in the subsequent analysis. Let the ratios of densities, viscosities, and fluid depths be denoted by $m=\mu_2/\mu_1$, $r=\rho_2/\rho_1$, and $n=d_2/d_1$, respectively, where subscripts 1 and 2 will denote quantities in the air and oil layers, respectively. For reference, $m\approx 5555$ and $n\approx100$ for 100cSt silicone oil and depths typical in experiments. 
We define the Reynolds number characterizing the air flow in the lubrication layer as $\mathrm{Re}= \rho_1 U_{\mathrm{w}} d_1/\mu_1$
and that characterizing the air flow around the droplet as $\mathrm{Re}_{\mathrm{d}}=\mathrm{Re}R/d_1=\rho_{1}U_{\mathrm{w}}R/\mu_{1}$. We also define a non-dimensional surface tension, an inverse Weber number, by $S\equiv \sigma/\left(\rho_1 d_1 U_{\mathrm{w}}^2\right)$. 

We non-dimensionalize the flow velocity by the wall velocity so that $U=U_{\mathrm{dim}}/U_{\mathrm{w}}$, where $U_{\mathrm{dim}}$ is the dimensional base flow. The non-dimensional pressure is given by $p=p_{\mathrm{dim}}/(\rho_1 U_{\mathrm{w}}^2)$. Lengths are non-dimensionalized with respect to the thin air layer so that $x=x_{\mathrm{dim}}/d_1 \; ,y=y_{\mathrm{dim}}/d_1$. We will proceed to work exclusively with non-dimensional variables.

The base flow in the upper and lower regions is given by
\begin{equation}\label{eq:baseflow}
U(y)= \left\{\begin{array}{ll}
a_1 y +b, \; y\in(0,1) \\
a_2 y +b,\; y\in(-n,0),
\end{array}\color{white}\right\}
\end{equation}
where,
\begin{equation}
a_1=m/(m+n);\;a_2=1/(m+n);\;b=n/(m+n)
  \label{sy}.
\end{equation}

Let us define the total flow as $\boldsymbol{u}=(U(y),0)+\boldsymbol{u}'$, where $\boldsymbol{u}'=(u,v)$ defines the perturbation velocity. We now consider the stability of small two-dimensional perturbations to the film. When assessing stability of parallel shear flows in a homogeneous medium, \citet{squire1933stability} proved that it is sufficient to consider two-dimensional disturbances. \citet{yih1955stability} extended this proof to the present case of non-uniform density and viscosity. We thus proceed to consider the evolution of Tollmien--Schlichting perturbations following the non-dimensionalization employed by \citet{yih1967instability}.

Consider two-dimensional wave disturbances (Tollmien--Schlichting waves) parameterized by a streamfunction, $\psi$, and perturbation pressure, $p$, as follows,
\begin{equation} \label{eq:pert_def}
\{\psi,p\}=\{\phi(y),f(y)\}\exp{\mathrm{i}\alpha(x-ct)},
\end{equation}
where $(u,v)=(\psi_y,-\psi_x)$ and subscripts denote partial differentiation. The non-dimensional wavenumber is related to the dimensional wavenumber $k$ by the relation, $\alpha=kd_1$. After plugging $\boldsymbol{u}=(U(y),0)+\boldsymbol{u}'$ into the Navier-Stokes equations, and using (\ref{eq:pert_def}) to eliminate pressure, we linearize to reach the standard Orr-Sommerfeld equation governing $\phi$,
\begin{equation}\label{eq:orrsom}
\phi^{\mathrm{iv}}-2\alpha^2\phi''+\alpha^4 \phi=\mathrm{i}\alpha \mathrm{Re}_i\left( (U(y)-c)(\phi''-\alpha^2\phi)-U''(y)\phi  \right),
\end{equation}
where $i=1$ in the air layer, $i=2$ in the oil layer, and $c$ is an eigenvalue to be determined by solving (\ref{eq:orrsom}) subject to the appropriate boundary conditions. We note that $U''(y)=0$ in the case of the base flow (\ref{eq:baseflow}) and $\phi$ must satisfy (\ref{eq:orrsom}) in the bulk, for $y\in (-d_2,0) \cup (0,d_1)$. It is thus convenient to express $\phi$ in terms of the functions $\Phi$ and $\Psi$, defined in the air and oil respectively, such that 
\begin{equation}
\phi(y)= \left\{ \begin{array}{ll}
\Phi(y), y\in (0,1) \\
\Psi(y), y\in (-n,0)\;, \\
 \end{array}\color{white}\right\}
  \label{piecewise_phi}
\end{equation}
with $\Phi$ and $\Psi$ both satisfying (\ref{eq:orrsom})
\begin{equation}\label{eq:uplow_orrsomfull}
 \begin{array}{ll}
\Phi^{\mathrm{iv}}-2\alpha^2\Phi''+\alpha^4 \Phi=\mathrm{i}\alpha \mathrm{Re} (U(y)-c)(\Phi''-\alpha^2\Phi)\\
\Psi^{\mathrm{iv}}-2\alpha^2\Psi''+\alpha^4 \Psi=\mathrm{i}\alpha \frac{r \mathrm{Re}}{m} (U(y)-c)(\Psi''-\alpha^2\Psi).\\
 \end{array}
\end{equation}

At the boundaries $y=-d_2$ and $y=d_1$, the flow must satisfy no-slip conditions. Since the base flow satisfies the no-slip conditions, so too must the perturbations. Thus,
\begin{equation}
\Phi_y(1)=0;\;
\Phi(1)=0;\;
\Psi_y(-n)=0;\;
\Psi(-n)=0.
  \label{eq:fullnoslip}
\end{equation}
Across the interface, the velocity must be continuous. The linearized interface height, $\eta(x,t)$, is defined by the kinematic condition,
\begin{equation}
    \left( \frac{\partial }{\partial t}+U\frac{\partial }{\partial x}\right)\eta=v=-\mathrm{i} \alpha \exp{\mathrm{i}\alpha(x-ct)},
\end{equation}
so that, after defining $c' \equiv c-U(0)$, the interface is given by
\begin{equation}\label{eq:etadef}
    \eta=\frac{\phi(0)}{c'}\exp{\mathrm{i}\alpha(x-ct)}.
\end{equation}
The continuity of vertical velocity, at $y=0$, simply yields
\begin{equation}\label{eq:full_Vcont}
    \Phi(0)=\Psi(0).
\end{equation}
However, the continuity of horizontal velocity must be enforced at the interface defined in (\ref{eq:etadef}). A Taylor expansion for small interface perturbations then yields the horizontal velocity continuity condition
\begin{equation}\label{eq:fullhorcont}
    \Phi'(0)-\Psi'(0)=\frac{\Phi(0)}{c'}\left( a_2-a_1 \right).
\end{equation}
The continuity of shear stress at the interface takes the form
\begin{equation}\label{fullshearcont}
    \Phi''(0)+\alpha^2\Phi(0)=m\left(\Psi''(0)+\alpha^2\Psi(0)\right).
\end{equation}
Finally, the continuity of normal stress may be written as
\begin{equation}
\begin{array}{ll}
-\mathrm{i}\alpha \mathrm{Re}\left(c'\Phi'  +  a_1\Phi\right)& -  (\Phi'''-\alpha^2\Phi')+ 2  \alpha^2\Phi' + \\
\mathrm{i}r\alpha\mathrm{Re}(c'\Psi'+a_2\Psi)  &+m(\Psi'''-\alpha^2\Psi')-2\alpha^2m\Psi'\\
 &=  \mathrm{i} \mathrm{Re}\alpha^3 S\Phi /c',
\end{array}\label{eq:fullnormshear}
\end{equation}
where all quantities are evaluated at $y=0$. The eigenvalue problem for $c$ is now fully specified through equations (\ref{eq:uplow_orrsomfull})-(\ref{eq:fullnormshear}). In order to make progress in determining the stability of various wavenumber perturbations, $\alpha$, we note that in the experiments of interest, the nondimensional wavenumber is small with $\alpha=2\pi d_1/\lambda \sim 2\pi (2\mu\mathrm{m}) /(\mathrm{1 mm})\approx 0.03$. Therefore, we are justified in confining our attention to perturbations with $\alpha \ll 1$. In the following section, we solve for $\Phi$ and $\Psi$ order by order in the small parameter $\alpha$ via a perturbation analysis.

\subsection{Long wavelength perturbation expansion}\label{perturb}
To solve the system of equations of \S \ref{mathform} perturbatively, we begin by writing $\Phi(y)=\Phi_0(y)+\alpha \Phi_1(y)+\alpha^2\Phi_2(y)+\alpha^3\Phi_3(y)+\mathcal{O}\left(\alpha^4\right)\;$ and $\Psi(y)=\Psi_0(y)+\alpha \Psi_1(y)+\alpha^2\Psi_2(y)+\alpha^3\Psi_3(y)+ \mathcal{O}\left(\alpha^4\right)\;$. We also write $c'=c_0'+\alpha c_1+\alpha^2 c_2+\alpha^3 c_3+\mathcal{O}\left(\alpha^4\right)\;$. Plugging these expressions into (\ref{eq:orrsom} -\ref{eq:fullnormshear}), we obtain a hierarchy of equations in increasing powers of $\alpha$ that can be solved sequentially. At each order, an eigenvalue problem emerges for the corresponding wave speed correction; if complex, these corrections correspond to temporal growth or decay. It will become clear that we need to solve to $\mathcal{O}(\alpha^3)$ to capture the stabiliting effect of surface tension, since the pre-factor of $S$ in (\ref{eq:fullnormshear}) is $\mathcal{O}\left(\alpha^3\right)$.

\subsubsection{$\mathcal{O}(1)$ Equations}\label{perturb0}
The zeroth order form of (\ref{eq:uplow_orrsomfull}) simplifies to
\begin{equation}\label{eq:uplow_orrsom0}
\Phi_0^{\mathrm{iv}}=0;\;\;
\Psi_0^{\mathrm{iv}}=0,
\end{equation}
whose general solution is simply a set of third degree polynomials for $\Phi_0(y)$ and $\Psi_0(y)$. We are free to set the constant term to unity, since the resulting eigenvalue problem will only be specified up to a constant scaling. We can then write $\Phi_0=1+A_1 y+A_2 y^2 + A_3 y^3$ and $\Psi_0=1+B_1 y+B_2 y^2 + B_3 y^3$. With this choice, the vertical velocity continuity (\ref{eq:full_Vcont}) is automatically upheld so that
\begin{equation}\label{eq:Vcontorderbyorder}
    \Phi_0(0)=\Psi_0(0)=1.
\end{equation}
Six undetermined coefficients (three in each layer) remain undetermined, along with the wave speed, all to be determined by application of the following boundary conditions. 

The four homogeneous wall conditions (\ref{eq:fullnoslip}) must be satisfied by $\Phi_i$ and $\Psi_i$ at each order of $\alpha$. 
At $\mathcal{O}(1)$, the horizontal velocity continuity (\ref{eq:fullhorcont}) becomes
\begin{equation}\label{eq:horcont0}
    \Phi_0'(0)-\Psi_0'(0)=\frac{\Phi_0(0)}{c'_0}\left( a_2-a_1 \right).
\end{equation}
The continuity of shear stress at the interface then simplifies to,
\begin{equation}\label{eq:shearcont0}
    \Phi''_0(0)=m\Psi''_0(0).
\end{equation}
Finally, the continuity of normal stress may be written as,
\begin{equation}
\begin{array}{ll}
 -  \Phi_0'''+ m\Psi_0'''=  0,
\end{array}\label{eq:normshear0}
\end{equation}
where all quantities are evaluated at $y=0$. The problem for $c_0'$ is now fully specified, through the following seven equations: the four equations in (\ref{eq:fullnoslip}), along with (\ref{eq:horcont0}), (\ref{eq:shearcont0}), and (\ref{eq:normshear0}). The seven unknowns are the three undetermined polynomial coefficients of each of $\Phi_0$ and $\Psi_0$, and the zeroth order wave speed $c_0'$. The linear algebra problem for the coefficients can be solved symbolically in Mathematica or numerically in MATLAB. We implemented both approaches and cross-checked results at all orders and find the approaches to be in agreement. \citet{yih1967instability} also solved the system up to $\mathcal{O}(\alpha)$; as a third check, we compared our results to his and found favourable agreement. We proceed to outline the $\mathcal{O}(\alpha)$, $\mathcal{O}(\alpha^2)$ and $\mathcal{O}(\alpha^3)$ equations. We suppress the symbolic solutions for the wave speed at each order for brevity, and report the final results of the perturbation analysis in the predictions given in figure \ref{fig:results}.

\subsubsection{$\mathcal{O}(\alpha)$ Equations}\label{perturb1}
The $\mathcal{O}(\alpha)$ form of (\ref{eq:uplow_orrsomfull}) simplifies to
\begin{equation}\label{eq:uplow_orrsom1}
 \begin{array}{ll}
\Phi_1^{\mathrm{iv}}=\mathrm{i} \mathrm{Re} (a_1 y -c_0')\Phi_0''\\
\Psi_1^{\mathrm{iv}}=\mathrm{i} \frac{r \mathrm{Re}}{m} (a_2 y -c_0')\Psi_0''.\\
 \end{array}
\end{equation}
whose homogeneous solution is again a set of third degree polynomials for $\Phi_1(y)$ and $\Psi_1(y)$. The particular solution is obtained through the direct integration of the zeroth order solution (cubic polynomial) as obtained in \S \ref{perturb0}. All constants of integration can be set to zero since this corresponds to the homogeneous part of the solution. We can then write the homogeneous part of the solution as $\Phi_1=1+A_1^{(1)} y+A_2^{(1)} y^2 + A_3^{(1)} y^3$ and $\Psi_1=1+B_1^{(1)} y+B_2^{(1)} y^2 + B_3^{(1)} y^3$, where we have once again set the constant term to unity. Again, and at all subsequent orders, this choice ensures the vertical velocity continuity (\ref{eq:full_Vcont}) is automatically satisfied. The six undetermined coefficients and the wave speed correction $c_1$ are determined by the boundary conditions. 

The four homogeneous wall conditions (\ref{eq:fullnoslip}) must be satisfied by $\Phi_i$ and $\Psi_i$ at each order of $\alpha$. 
At this order, the horizontal velocity continuity (\ref{eq:fullhorcont}) becomes
\begin{equation}
\begin{array}{ll}
    \Phi_1'(0)-\Psi_1'(0)=\frac{-c_1\Phi_0(0)+c_0'\Phi_1(0)}{(c_{0}')^{2}}\left( a_2-a_1\right).
\end{array}
\end{equation}
The continuity of shear stress again simplifies to the form,
\begin{equation}
    \Phi''_1(0)=m\Psi''_1(0).
\end{equation}
Lastly, the continuity of normal stress may be written as,
\begin{equation}
\begin{array}{ll}
-\mathrm{i} \mathrm{Re}\left(c'\Phi_0'  +  a_1\Phi_0\right)  -  \Phi_1''' + \mathrm{i}r\mathrm{Re}(c'\Psi_0'+a_2\Psi_0) +m\Psi_1''' =  0,
\end{array}\label{eq:normshearOa}
\end{equation}
where all quantities are evaluated at $y=0$. The problem for $c_1$, along with the six undetermined coefficients, is now fully specified. The simple linear algebra problem can again be solved symbolically in Mathematica or numerically in MATLAB. We now move to the $\mathcal{O}(\alpha^2)$ and $\mathcal{O}(\alpha^3)$ equations. At this point, we mention that $c_1$ is purely complex, as can be seen by examination of the perturbation equations at $\mathcal{O\left(\alpha\right)}$. Thus, this correction corresponds to the growth rate of the perturbation. As the corrections to the wave speed alternate between being purely real and purely imaginary, it is required to solve to $\mathcal{O}(\alpha^3)$ to get the next imaginary correction to the growth rate. This is consistent with the fact that the effects of surface tension, which should be stabilizing, first enter at that order.

\subsubsection{$\mathcal{O}(\alpha^2)$ Equations}\label{perturb2}
The $\mathcal{O}(\alpha^2)$ form of (\ref{eq:uplow_orrsomfull}) simplifies to
\begin{equation}\label{eq:uplow_orrsom2}
 \begin{array}{ll}
\Phi_2^{\mathrm{iv}}-2\Phi_0''=\mathrm{i} \mathrm{Re} ((a_1 y -c_0')\Phi_1''-c_1\Phi_0'')\\
\Psi_2^{\mathrm{iv}}-2\Psi_0''=\mathrm{i} \frac{r \mathrm{Re}}{m} ((a_2 y -c_0')\Psi_1''-c_1\Psi_0'')).\\
 \end{array}
\end{equation}
We can then write the homogeneous part of the solution as $\Phi_2=1+A_1^{(2)} y+A_2^{(2)} y^2 + A_3^{(2)} y^3$ and $\Psi_2=1+B_1^{(2)} y+B_2^{(2)} y^2 + B_3^{(2)} y^3$, where have once again set the constant term to unity to ensure vertical velocity continuity. The particular solution is obtained through direct integration of the terms in (\ref{eq:uplow_orrsom2}) involving the zeroth and first order solutions. Integration constants are again set to zero. Six undetermined coefficients, and the wave speed correction $c_2$, are determined by application of the boundary conditions as follows. 

The four homogeneous wall conditions (\ref{eq:fullnoslip}) must be satisfied by $\Phi_2$ and $\Psi_2$. 
At this order, the horizontal velocity continuity (\ref{eq:fullhorcont}) becomes
\begin{equation}
\begin{array}{ll}
    \Phi_2'(0)-\Psi_2'(0)=\\
    \frac{c_1^2\Phi_0-c_0'c_2\Phi_0-c_0 c_1 \Phi_1 +(c_0')^2\Phi_2}{(c_{0}')^{3}}\left( a_2-a_1\right).
\end{array}
\end{equation}
The continuity of shear stress simplifies to the form
\begin{equation}
    \Phi''_2(0)+\Phi_0(0)=m\left(\Psi''_2(0)+\Psi_0(0)\right).
\end{equation}
Finally, at this order, the continuity of normal stress may be written as,
\begin{equation}
\begin{array}{ll}
3\Phi_0'-  \mathrm{i}\mathrm{Re}\left(a_1\Phi_1+c_0'\Phi_1'+c_1\Phi_0'\right)+\\
\mathrm{i}r\mathrm{Re}\left(a_2\Psi_1+c_0'\Phi_1'+c_1\Phi_0'\right)-
\Phi_2'''+m(\Psi_2'''-3\Psi_0')=0,
\end{array}\label{eq:pertnormshear2}
\end{equation}
where all quantities are evaluated at $y=0$.

The problem for $c_2$, along with the six undetermined coefficients, is now fully specified. The linear algebra problem can again be solved symbolically in Mathematica or numerically in MATLAB. We now outline the $\mathcal{O}(\alpha^3)$ equations and then compare our theoretical results to experiment.

\subsubsection{$\mathcal{O}(\alpha^3)$ Equations}\label{perturb3}
The $\mathcal{O}(\alpha^3)$ form of (\ref{eq:uplow_orrsomfull}) simplifies to
\begin{equation}\label{eq:uplow_orrsom3}
 \begin{array}{ll}
\Phi_3^{\mathrm{iv}}-2\Phi_1''=\mathrm{i} \mathrm{Re} ((a_1 y -c_0')\left(-\Phi_0+\Phi_2''\right)-c_1\Phi_1''-c_2\Phi_0'')\\
\Psi_3^{\mathrm{iv}}-2\Psi_1''=\mathrm{i} \mathrm{Re}\frac{r}{m} ((a_2 y -c_0')\left(-\Psi_0+\Psi_2''\right)-c_1\Psi_1''-c_2\Psi_0'').\\
 \end{array}
\end{equation}
By writing a general homogeneous solution in each layer, and integrating (\ref{eq:uplow_orrsom3}) to obtain a particular solution, we again reach six undetermined coefficients, and the wave speed correction $c_3$, to be determined via application of the following boundary conditions. 

The four homogeneous wall conditions (\ref{eq:fullnoslip}) must be satisfied by $\Phi_3$ and $\Psi_3$. 
At this order, the horizontal velocity continuity (\ref{eq:fullhorcont}) becomes
\begin{equation}
\begin{array}{ll}
    \Phi_3'(0)-\Psi_3'(0)=\\
    \frac{\left(-c_1^3+2c_0'c_1c_2-(c_0')^2c_3\right)\Phi_0+c_0'(c_1^2-c_0'c_2)\Phi_1-(c_0')^2c_1\Phi_2+(c_0')^3\Phi_3}{(c_0')^4}
\end{array}
\end{equation}
The continuity of shear stress simplifies to the form,
\begin{equation}
\Phi''_3(0)+\Phi_3(0)=m\left(\Psi''_3(0)+\Psi_1(0)\right).
\end{equation}

Finally, at this order, the continuity of normal stress may be written as,
\begin{equation}
\begin{array}{ll}
\mathrm{i}\mathrm{Re}\Big(
\frac{S}{c_0'}\Phi_0+a_1\Phi_2+c_2\Phi_0'+  c_1\Phi_1'+c_0'\Phi_2'-\\
r(a_2\Psi_2+c_2\Psi_0'+c_1\Psi_1'+c_0'\Psi_2')
\Big)  =3\Phi_1'-\Phi_3'''+m(\Psi_3'''-3\Psi_1')
\end{array}\label{eq:pertnormshear3}
\end{equation}
where all quantities are evaluated at $y=0$. The linear algebra problem for $c_3$ and the undetermined coefficients is again solved symbolically in Mathematica.

\subsection{The Fastest Growing Mode}\label{FGM}
With $c$ known, the phase speed of a given wave is given by $\mathrm{Re}\{c\}=c_0+\alpha^2 c_2$. Moreover, the growth rate of the wave is given by $G=\alpha \mathrm{Im}\{c\}=\alpha^2 c_1+\alpha^4 c_3$ (see (\ref{eq:etadef})). The fastest growing mode is then given by solving $G'(\alpha_{\mathrm{max}})=0$, so that the fastest growing wavenumber is given by 
\begin{equation}\label{eq:wavenumber_exp}
\alpha_{\mathrm{max}}=\sqrt{-\frac{c_1}{2 c_3}}
\end{equation}
This result is valid provided the computed wavenumber conforms to our long-wavelength assumption, so that $\alpha_{\mathrm{max}}\ll 1$; this assumption requires $c_3 \gg c_1$, which we find to be true, as seen in figure \ref{fig:results}. The fastest growing mode is selected by surface tension since $c_3$ depends on $S$. Redimensionalizing, the fastest growing wavelength is
\begin{equation}
\lambda_{\mathrm{max}}=2\sqrt{2}\pi d_1\sqrt{-\frac{c_3}{c_1}}.
\end{equation}

Generally, if there is instability, we have that $c_1>0$, corresponding to the viscous shear instability that was observed by \citet{yih1967instability}. We note that $c_3<0$, indicating the stabilizing effect of surface tension arising at $\mathcal{O}(\alpha^3)$. The non-dimensional wave phase speed, corresponding to the fastest growing mode, is then simply given by $c_{\mathrm{max}}=c_0+\alpha^2 c_2=c_0'+b+\alpha^2 c_2$.

\section{Determination of $d_2$ and Comparison with Experiments}\label{detd2}
\begin{figure}
\centering
 \includegraphics[width=1\linewidth]{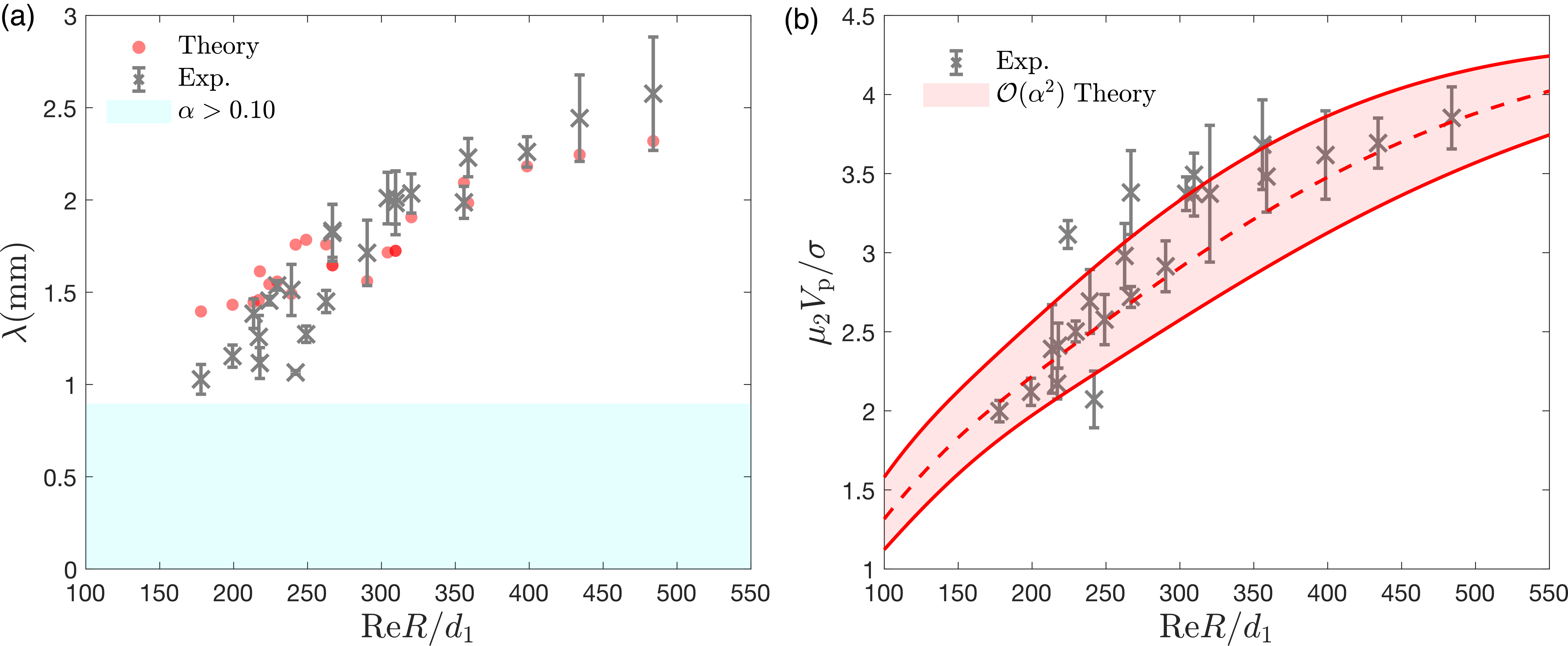}
 \caption{\label{fig:results} Comparison of the predictions from the perturbation analysis of \S \ref{mathform} with experiments. a) Red dots give the fastest growing wavelength as predicted by our model. The lubrication layer thickness is given by the scaling $d_1=0.34H\mathrm{Ca}^{\frac{2}{3}}$, while $d_2$ is the effective shear length in the drop, prescribed by the scaling $d_2=A H$, with $A=0.08$ representing the only fitting parameter in the model. The region where $\alpha>0.1$ is shaded, to illustrate the range of validity of our long wavelength approximation, $\alpha \ll 1$. b) Red lines give theoretical predictions of the wave speed according to our theory to $\mathcal{O}(\alpha^2)$ plotted versus the drop Reynolds number, $\mathrm{Re}_{\mathrm{d}}=\rho_1 U_{\mathrm{w}}R/\mu_1$. The dotted line is computed using the average radius of the droplets across all experiments. The upper and lower solid red lines represent theoretical predictions for radius values one standard deviation above and below this mean, respectively.}
\end{figure}

We now fit $d_2$, the only free parameter in our model, which defines the effective shear layer thickness in the droplet. As described in \S \ref{physicalpic}, we assume the scaling $d_2=AH$, where $A$ is to be determined. We determine $A$ by fitting to the wavelength data across our experiments. Taking $A=0.08$, we achieve the theoretical predictions to the data shown in \ref{fig:results}(a). 

Experimental phase speeds are plotted in figure \ref{fig:results}(b). Our theory at $\mathcal{O}(\alpha^2)$ is presented in red lines, yielding fair agreement with experiments. The vertical spread in the model prediction at $\mathcal{O}(\alpha^2)$ comes from a weak dependence of the theory on variations of dimensionless groups, other than the Reynolds number, between experiments. The dotted line represents the theoretical prediction with the drop radius fixed at the average value of the drops across all experiments, $R_{\mathrm{avg.}}=2.57 \mathrm{mm}$. Then $U$ is varied, for this fixed radius, in order to attain all values of $\mathrm{Re}$ reported in the plot. The upper and lower solid red lines are theoretical predictions corresponding to the radii $R- \sigma_R$  and $R+ \sigma_R$, respectively, where $\sigma_R=0.49 \mathrm{mm}$ is the standard deviation of the spread of radii across all experiments. Overall, there is a fair agreement between the theory and experiment. 

\section{Discussion and Conclusion}\label{disc}
We have examined a propagating wave instability that develops beneath a levitating droplet as reported by \citet{sawaguchi2019droplet}. By making assumptions consistent with experimental observations, we have developed a simplified model of the lubrication flow beneath the droplet. We then analysed the stability of the idealized base state to small perturbations of the Tollmien-Schlichting type. Our analysis builds on the work of \citet{yih1967instability}, who examined a long wavelength instability due to viscosity stratification. Notably, \citet{yih1967instability} considered a small wavenumber ($\alpha \ll 1$) expansion and solved to first order in $\alpha$ to assess the onset of instability.
Conversely, we were obliged to solve to third order in $\alpha$ in order to capture the effect of surface tension. We find that surface tension selects a most unstable mode, whose wavelength and phase speed are in adequate agreement with available experimental data.

The main inputs to our model are $d_1$, the thickness of the air film beneath the drop, and $d_2$, an effective shear layer thickness inside the droplet. The former input, $d_1$, is governed by the scaling presented by \citet{gauthier2016aerodynamic}. The only fitting parameter in our model is thus $A=d_2/H\approx 0.08$, which was found by
fitting to our experimental data for the wavelength of instabilty indicated as red points in figure \ref{fig:results}(a). With $A \approx 0.08$ fixed, the theory predicts the phase speed of waves evident in figure \ref{fig:results}(b). The data trend of phase speed increasing with the drop Reynolds number is captured well by the theory. 
Note that the small vertical spread in theoretical predictions evident in \ref{fig:results}(b) indicates that the phase speed depends weakly on parameters other than the drop Reynolds number, $\mathrm{Re}_{\mathrm{d}}$, including the drop radius. While it would have been preferable to plot the data and theory in Fig. \ref{fig:results} against the lubrication Reynolds number, as the lubrication thickness $d_1$ was not measured directly in experiments, we were obliged to plot against the droplet Reynolds number, $\mathrm{Re}_{\mathrm{d}}$.

While our analysis applies to the idealized geometry of figure \ref{fig:FlowSchem}(b), it only serves as an approximate model for the levitating droplet under consideration. We believe the slight discrepancy between theory and experiment, evident in figure \ref{fig:results}(b), is due to both geometric simplifications (see \S \ref{physicalpic}) as well as our assumed scaling for $d_2$. For example, weak curvature of the drop's lower surface \citep{lhuissier2013levitation} might augment phase velocities relative to those on a flat interface. Likewise, three dimensional aspects of the flow reported by \citet[Supplementary Movie 3]{sawaguchi2019droplet} suggest that our assumption of purely unidirectional flow may have limited validity. Finally, one does not expect our inferred shear layer scaling, $d_2 \sim 0.08 H$, to remain valid outside the experimental regime considered in the present paper.

Overall, we have found adequate agreement between our idealized theoretical model and experimental observations. We thus believe that we have elucidated the key mechanism of instability. Specifically, the waves observed by \citet{sawaguchi2019droplet} emerge due to a viscous shear instability of the lubricating air flow beneath the droplet. While the problem involves a variety of dimensionless groups, the unstable mode's wavelength and phase speed depend primarily on the Reynolds number. We conclude by noting that similar wave instabilities might appear in other systems with high interfacial shear, including the rolling drop system of \citet{gauthier2016aerodynamic}, as well as other high-speed noncoalescence events~\cite{wad,thrasher2007bouncing,Kolinski_2014,kavehpour2015coalescence}. Similar instabilities might also appear in inverse Leidenfrost systems \cite{gauthier2019self} with high viscosity drops.

\noindent \textbf{Funding.} K.M was generously funded by a Mathworks fellowship during this work. JB thanks the NSF for financial support through grant CMMI-2154151.

\begin{acknowledgments}
The authors thank Ayumi Matsuda for help in obtaining the image in Fig. \ref{fig:fig_exp_im}. The authors are grateful for valuable discussions with T.R. Akylas, and K.J. Burns.
\end{acknowledgments}

\appendix

\section{Inviscid and Viscus Potential Flow Models}\label{Inv_Visc_pot_flos}
At first glance, the waves appear to be of the classic Kelvin--Helmholtz (KH) type: the air lubrication layer has velocities near 1-2 m/s while the drop is almost at rest, giving rise to a shear flow. We now crudely approximate the flow in the potential flow limit and show the analysis to be lacking through comparison with experimental data. We first consider an inviscid potential flow model using parameter values typical in the experiments of \citet{sawaguchi2019droplet}.

Considering an air layer with thickness 10 $\mu$m that moves at velocity 2 m/s and a liquid layer of 100 cSt silicone oil of thickness 1 mm at rest, and taking the surface tension value $\sigma = 21 \mathrm{mN/m}$, inviscid theory predicts maximum growth of the wavelength $\lambda_{\mathrm{i}} = 2.14$ mm with phase speed $V_{\mathrm{i}}=0.087 \mathrm{m/s}$. While the predicted wavelength is within a factor of 2 of the experimental value, $\lambda_{\mathrm{exp}}=1.3\pm0.1 $mm, the phase speed is far below the observed value of $V_{\mathrm{exp}}\approx0.72\mathrm{m/s}$. 
Considering now the viscous potential flow framework of \citet{funada2001viscous}, where viscous effects are included only in the normal stress interfacial conditions, the wavelength of maximum growth is worse,  $\lambda_{\mathrm{vis}}=3.59\mathrm{mm}$, while the wave speed is still underestimated with $V_{\mathrm{vis}}=0.095\mathrm{m/s}$. It is thus necessary to incorporate the effects of shear stress to obtain predictions for the wavelength and phase speed.

\section{Analysis of Natural Vibration Modes}\label{resjust}
We here demonstrate that the frequency of the observed waves is an order of magnitude different from the first natural vibration modes of the droplet. 

The upper portion of the droplet is approximately spherical. It should thus vibrate naturally roughly according to the inviscid natual oscillation formula given by \citet[\S 253]{Lamb1975}, where the $n^{th}$ mode of vibration has a frequency of 
\begin{equation}\label{eq:vibs}
\omega_n = \sqrt{n(n-1)(n+2)\frac{\sigma}{ \rho  R^3} }.
\end{equation}
In the smallest drops in our experiments ($R=1.7 \mathrm{mm}$), the first three vibrational modes according to (\ref{eq:vibs}) are given by $\omega_2=188/\mathrm{s}$, $\omega_3=364/\mathrm{s}$, $\omega_4=563/\mathrm{s}$. Note that droplet non-sphericity, as well as viscosity, have been shown experimentally to decrease these frequencies further \citep{perez1999oscillation}.

By way of comparison, the frequency of the waves observed beneath drops, as measured by counting the number of wave crests passing a given point per unit time is $\omega_{\mathrm{exp}}=2772/\mathrm{s}$. The experimental frequency was computed by observing $13$ wave crests passing by a given point over a video of length $30 \mathrm{ms}$.

The experimental value corresponds roughly to the $n=12$ mode of the smallest drop. For larger drops, the value of $n$ for which $\omega_n=2775/\mathrm{s}$ would be even higher. Due to the increased damping experienced by higher modes, it is unlikely that the excitation of such high-$n$ modes is responsible for the observed waves. 


\bibliography{manuscript_text}

\end{document}